\documentclass[11pt]{article}
\usepackage[utf8]{inputenc}
\usepackage{sgame}
\usepackage{amssymb}
\usepackage{amsmath}
\usepackage{graphicx}
\usepackage{color}
\usepackage{float}
\usepackage{tikz}
\usepackage[letterpaper,pdftex]{geometry}
\usepackage{longtable}
\usetikzlibrary{arrows,automata,fit,scopes,calc,matrix,positioning,decorations.pathmorphing,decorations.pathreplacing}
\usepackage[all]{xy}
\newtheorem{proposition}{Proposition}
\newtheorem{example}{Example}
\usepackage[sc]{mathpazo}
\usepackage[T1]{fontenc}
\newtheorem{defi}{Definition}
\newtheorem{teo}{Theorem}

\newtheorem{propi}{Proposition}

\title{Dynamic Arrangements in Economic Theory: Level-Agnostic Representations}
\author{Fernando Tohm\'e\\
Departamento de Econom\'{\i}a - Universidad Nacional del Sur\\
Instituto de Matem\'atica de Bah\'{\i}a Blanca - CONICET
}
\date{}

\begin{document}
\maketitle

\maketitle
\begin{abstract}
    If Economics is understood as the study of the interactions among intentional agents, being rationality the main source of intentional behavior, the mathematical tools that it requires must be extended to capture systemic effects. Here we choose an alternative toolbox based on Category Theory. We examine potential {\em level-agnostic} formalisms, presenting three categories, $\mathcal{PR}$, $\mathcal{G}$ and an encompassing one, $\mathcal{PR-G}$. The latter allows for representing dynamic rearrangements of the interactions among different agents.
\end{abstract}
\section{Introduction}

\noindent It is hard to define with precision the actual scope of Economics. Perhaps the best-known definition was given by Lionel Robbins (1932): 
\begin{quote}
{\it Economics is the science which
studies human behavior as a relationship between ends and
scarce means which have alternative uses.}
\end{quote}

\noindent While widely accepted, this characterization is unsatisfactory in many ways. In particular, for not taking into account crucial developments that reshaped the discipline in the last nine decades.\\

\noindent Accordingly, a more general definition could be 
\begin{quote}
{\it Economics studies the {\em interaction} among {\em intentional} entities.} 
\end{quote}

\noindent This summarizes most if not all the research activities of contemporary economists. The term ``entity'', which is introduced to refer to firms, institutions, and other non-human economic agents, covers the extension of economic analyses to all kinds of things able to exhibit agency, ranging from social groups to robots.\footnote{No wonder that other social scientists think that Economics is an ``imperialist'' discipline!}\\

\noindent This extended notion of Economics, has been formalized in a rather narrower sense, using tools ranging from Calculus and Linear Algebra to Functional Analysis and Algebraic Topology. Modern Economic Theory as well as a good deal of Econometrics have been shaped using methods drawn from those fields. But the full meaning of the alternative characterization given above can only be captured by conceiving Economics as a system composed of other systems. While contemporary disciplines like Computer Science have fully embraced this view, economists have been reluctant to adopt it.\\

\noindent In this contribution we explore possible formalisms that may support the development of tools for an extended conception of Economics. While this is a wide-ranging project, we consider here two issues:

\begin{itemize}
\item How to deal with the complications inherent in attempts to sever different ``local'' interactions as if all the others
remained fixed.
\item How to scale up the solutions with the aggregation of the problems of interest.
\end{itemize}

\noindent Both issues reveal the need for a {\em level-agnostic} (or {\em continuous with respect to subagents}) Economic Theory. This paper lays the ground for a such model. We start by noting that Economics has a well-defined notion of agent defined in terms of a given preference relation over the space of alternatives. Then, the agent is said {\it rational} if she chooses the most preferred alternatives among those that are feasible for her.\\

\noindent In applications, it is customary to reduce the analysis to a subspace of the space of alternatives, simplifying the problem of making a decision. But this comes at the price of assuming the independence of the preferences
over the subspace from the preferences over the rest of the
larger space of alternatives.\\

\noindent In this initial version we first present a way of ensuring the consistency of the solutions found for the different subspaces. Then, another approach to the coordination of independent context is given, in this case involving games with shared players.\\

\noindent The final part of this paper presents a generalization, integrating both models, in which interactions are no longer fixed, but can evolve according to the inputs and outputs. In this as well as in the previous two models we apply the mathematical framework of Category Theory (\cite{Tohme1}).

\section{Mathematical Preliminaries}

\noindent As is well-known, Category Theory has provided a framework without which most of the contemporary results in both Algebraic Geometry and Topology would not have been found \cite{AT}. As repeatedly shown in actual mathematical practice, the language of Set Theory remains insufficient for capturing perspicuously the nuances prevalent in those fields \cite{Marquis}. One reason is that unlike Set Theory the categorical approach allows for both the maximization of the ``external'' scope of its formal results {\em and} the controlled ``internal'' sensitivity to particular differences in content within the representation of mathematical structures. While Category Theory might thereby also seem to be a natural choice of a formal language for representing the decision-making problems outlined above, we have to note that Economics has been reluctant to adopt it.\footnote{Some notable exceptions are \cite{Ghani}, \cite{Jules}, \cite{Winschel} and \cite{Rozen}. In turn, \cite{Crespo} presents arguments for the adoption of the categorical language in Economics.}\\

\noindent In this paper we draw heavily on the literature on Category Theory, although our results are clearly elementary. We will now present the basic concepts that will be used in subsequent sections. For further details and clarification, see the excellent presentations of Goldblatt (\cite{Goldblatt}), Barr \& Wells (\cite{Barr}), Ad\'amek et al. (\cite{Adamek}), Lawvere and Shanuel (\cite{Lawvere}), Spivak (\cite{Spivak}), Fong and Spivak (\cite{Fong}), Southwell (\cite{South}) or Cheng (\cite{Cheng}).\\

\noindent A category $\mathbf{C}$ consists of a set of {\em objects}, $\mathbf{\mbox{Obj}}$ and a class of {\em morphisms} between pairs of objects. Given two objects $a, b \in \mathbf{\mbox{Obj}}$ a morphism $f$ between them is notated $f: a \rightarrow b$. Given another object $c$ and a morphism $g: b \rightarrow c$, we have that $f$ and $g$ can be composed, yielding $g \circ f: a \rightarrow c$ (COMPOSITION). Additionally, for every $a \in \mathbf{\mbox{Obj}}$, there exists an {\em identity} morphism, $\mbox{Id}_a : a\rightarrow a$. Morphisms are required to obey two rules: $(i)$ if $f: a \rightarrow b$, $f \circ \mbox{Id}_a = f$ and $\mbox{Id}_b \circ f = f$ (IDENTITY); $(ii)$ given $f: a \rightarrow b$, $g: b \rightarrow c$ and $h: c \rightarrow d$, $(h \circ g) \circ f = h \circ (g \circ f) : a \rightarrow d$ (ASSOCIATIVITY).\\

\noindent Examples of categories are $\mathbf{SET}$ (the objects are sets, and the morphisms are functions between sets), $\mathbf{TOP}$ (the objects are topological spaces and the morphisms continuous functions), $\mathbf{POrd}$ (the objects are preorders and the morphisms are order-preserving functions), $\mathbf{Vec}$ (the objects are vector spaces and the morphisms linear maps), etc.\\

\noindent The terseness of categories facilitates diagrammatic reasoning. A diagram in which nodes represent objects and arrows represent morphisms allows to establish properties of a category. Diagrams that {\em commute}, i.e. such that all different direct paths of morphisms with the same start and end nodes are identified (that is, compose to a common morphism), indicate relations similar to those that can be established by means of equations.\\

\noindent Some of the most interesting constructions that can defined in categories are {\em limits} and {\em colimits} (duals of limits). Any limit (or colimit) captures a {\em universal property} on a family of diagrams with the same basic shape. This basic shape is captured by a {\em cone}, that is, an object $a$ and a family of arrows $\{f_a^{b_j}: a \rightarrow b_j\}_{j \in \mathcal{J}}$, such that for any pair $j, l \in \mathcal{J}$, if there exists a morphism $\gamma_{jl}: b_j \rightarrow b_l$ we have that $\gamma_{jl} \circ f_a^{b_j} = f_a^{b_l}$ (see Figure~1). \\

\begin{figure}[h]~\label{commut}
 \begin{displaymath}
\xymatrix{
a \ar[drr]_{f_a^{b_l}}  \ar[rr]^{f_a^{b_j}} & & b_j \ar[d]^{\gamma_{jl} }& \\
&  & b_l}
\end{displaymath}
\caption{Commutative diagram}
\end{figure}
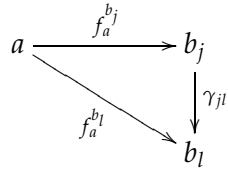

\noindent Then, given a class of cones of a given shape, a limit is an object $L$ in this class such that for every other cone $T$ in the class there exists a single morphism $T \rightarrow L$ such that the resulting combined diagram commutes. For instance, consider a family of cones of the shape depicted in Figure~2.

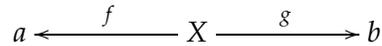
\begin{figure}[h]
 \begin{displaymath}
\xymatrix{
a && \ar[ll]_{f}  X  \ar[rr]^{g} & & b }
\end{displaymath}
\caption{The limit of cones of this shape defines the product  $a \times b$ }
\end{figure}

\noindent then, the limit is the {\em product} $a \times b$ and with arrows $p_1$ and $p_2$, the projections on the first ($a$) and second ($b$) components, respectively. For every other cone, with ``apex'' $X$ there is a unique morphism $!: X \rightarrow a \times b$ such that $f = p_1 \circ !$ and $g = p_2 \circ !$.\\

\noindent Examples of colimits are {\em direct sums} (in $\mathbf{\mbox{SET}}$, disjoint unions) and, somewhat confusingly called, {\em direct limits}, which in a self-contained description we will use to define {\em global solutions}.\\

\noindent Besides capturing interesting constructions common to many fields of Mathematics, Category Theory also provides tools for relating different categories to one another. This is achieved by means of mappings called {\em functors}. Given two categories $\mathbf{C}$ and $\mathbf{D}$ a functor $F$ from $\mathbf{C}$ to $\mathbf{D}$ maps objects from $\mathbf{C}$ into objects of $\mathbf{D}$ as well as arrows from the former to the latter category such that, if

$$ f: a \rightarrow b$$ 

\noindent in $\mathbf{C}$, then:

$$ F(f): F(a) \rightarrow F(b)$$

\noindent in $\mathbf{D}$. Furthermore $F(g \circ f) = F(g) \circ F(f)$ and $F(\mbox{Id}_a) = \mbox{Id}_{F(a)}$ for every object $a$ in $\mathbf{C}$.\\

\noindent These functors are called {\em covariant}. Another class, that of {\em contravariant} functors, is such that, if 

$$ f: a \rightarrow b$$ 

\noindent in $\mathbf{C}$, then:

$$ F(f): F(a) \leftarrow F(b)$$

\noindent in $\mathbf{D}$. Of particular interest are the contravariant functors $F: \mathbf{C} \rightarrow \mathbf{SET}$ (or a category of subsets of a given set), which are called {\em presheaves}. An intuitive interpretation is that given a morphism $a \rightarrow b$ in $\mathbf{C}$, the morphism $F(b) \rightarrow F(a)$ in $\mathbf{SET}$ is the {\em restriction} of the ``image'' under $F$ of $b$ over the ``image'' of $a$. Given an object $a$ in $\mathbf{C}$, $F(a)$ is called a {\em section} of $F$ over $a$. This can be extended to any family $B = \{b_j\}_{j \in  \mathcal{J}}$ of objects in $\mathbf{C}$: $F(B)$ is the section over $B$. In turn, given two families $B \subseteq B^{\prime}$ and the section over $B^{\prime}$, namely$F(B^{\prime})$ we can find its restriction over $B$, denoted $F(B^{\prime})_{|B}$, yielding $F(B)$.\\

\noindent Given a presheaf $F : \mathbf{C} \rightarrow \mathbf{SET}$, consider a class of objects $B$ in $\mathbf{C}$ and a cover $ \{K_j\}_{j \in  \mathcal{J}}$  (i.e. $B \subseteq \bigcup_{j \in  \mathcal{J}} K_j$). Let $\{k_j\}_{j \in  \mathcal{J}}$ be a sequence such that $k_j \in F(K_j)$ for each $j \in  \mathcal{J}$. The presheaf $F$ is said to be a {\em sheaf} if the following conditions are fulfilled:

\begin{itemize}
\item  {\em Locality}: For every pair $i,j  \in  \mathcal{J}$,  $k_{i|_{K_i \cap K_j}} = k_{j|_{K_i \cap K_j}}$ (i.e. the sections $a_i, a_j$ coincide over $V_i \cap V_j$), 
\item {\em Gluing}: There exists a unique $\bar{b} \in F(B)$ such that $\bar{b}_{|K_j} = k_j$ for each $j \in  \mathcal{J}$ (i.e. there exists a single object in the ``image'' of $B$ that when restricted to each set in the covering yields the section corresponding to that set).
\end{itemize}

\noindent This brief review of Category Theory provides the basic concepts necessary for the analysis to be carried out in the rest of the paper.\\

\noindent Other notions will be introduced in the following sections.


\section{Decision-making: Local vs. Global}

\noindent The traditional characterization of decision-making under certainty by an individual is as follows. Let ${\mathcal L}_i$ be a space of possible {\bf options} that an agent $i$ may select.\footnote{The meaning of these options depends on the context. If the agent is a consumer in a competitive market with a finite number of goods, she has to choose a vector of those commodities. In a planning problem, she has to select a plan specifying the amounts of resources used or consumed at each period of time.} Each $x \in {\mathcal L}_i$ is evaluated by means of a {\em utility} function, $U_i: {\mathcal L}_i \rightarrow \Re$. Given a family of constraints limiting the set of options open to the agent to $\hat{L}_i \subseteq{\mathcal L}_i$, the goal of the agent is to find some ${\mathbf x}^{*}$ that maximizes $U_i$ over $\hat{L}_i$. If we focus on the possible choices made by a single agent, we can drop the subindex $i$ from the notation for ${\mathcal L}$, $\hat{L}$ and $U$. We will reintroduce the dependence on the agents in the next sections to analyze the interaction between different agents.\\

\noindent In order to proceed, we first make some plausible assumptions. The space of options, ${\mathcal L}$, is presumed to be a (real) Hilbert space. That is, it is a complete metric space with an inner product.  Furthermore, in order to ensure the existence of a ${\mathbf x}^{*}$ we will also assume that $\hat{L}$ is a compact subset of ${\mathcal L}$, and that $U$ is a continuous function. Within this very general framework, it is then straightforward to induce a category-theoretical representation of the global optimization of $U$ over $\hat{L}$, that is, of ${\mathbf x}^{*}$ as a {\em direct limit}.\\

\noindent To begin, consider first a family $\{L^{k}\}_{k=0}^{\kappa}$ of closed linear subspaces of ${\mathcal L}$ and, for any given $k$, let us define the function

 \[\mbox{Proj}_k: {\mathcal L} \rightarrow \bigcup_{k=0}^{\kappa} L^{k}\] 
 
\noindent such that $\mbox{Proj}_k(x) = x^{k} \in L^{k}$, where $x^k$ is the {\em projection} of $x$ on $L^{k}$. The existence of such a projection is ensured by a straightforward application of the Linear Projection Theorem.\footnote{That is, $|x - x^k|= \min_{y \in L^{k}} |x -y|$, where $|\cdot|$ is the norm of ${\mathcal L}$.}\\

\noindent The projector operator $\mbox{Proj}_k$ will play a fundamental role in what follows. The intuition here is that we can think of each $L^k$ as the options set of a local problem. Therefore, the projection of a global solution ${\mathbf x}^{*}$ onto $L^k$ will return the point in $L^k$ which is the closest (i.e, the best!) to ${\mathbf x}^{*}$. Analogous approaches have been used successfully in several different contexts.\footnote{See, among others, Luenberger \cite{Luenberger}, where these methods are employed to model pricing assets whose payoffs are outside the span of marketed assets}.\\

\noindent In case the projection does not return a local solution, we can still define an operator, which we call $\Gamma_k: \hat{L} \rightarrow \hat{L}^k$ that formalizes the idea of a choice that is the closest to the projection (if it does not belong to the subspace):

$$\Gamma_k(x) = \{x^k \in \hat{{\mathbf X}}^{k} : x^k \in \mbox{argmin}_{y \in \hat{{\mathbf X}}^{k}} |y - \mbox{Proj}_k(x)| \}.$$

\noindent In some cases the global solution is not given, but must be sought by gluing together local ones ``prospectively'', in the hope of producing (or better, abducing) a consistent global result. In order to formalize this broadly abductive method for seeking a global solution, we need to take a second, slightly deeper plunge into category theory and start with the definition of a category of local problems.\\

\begin{defi}~\label{problem}
Let $\mathbf{\mathcal{PR}}$ be the category of local problems, where
\begin{itemize}
\item $\mbox{Obj}(\mathbf{\mathcal{PR}})$ is the class of objects. Each one, $s^k = \langle \hat{L}^{k}, u^{k}, \hat{{\mathbf X}}^{k} \rangle$ involves the maximization of the continuous utility function $u^{k}$ over the compact set $\hat{L}^{k} \subset L^{k}$, a closed linear subspace of ${\mathcal L}$, yielding a family of solutions $\hat{{\mathbf X}}^{k}$.
\item a morphism $\rho_{kj}: s^{k} \rightarrow s^{j}$ is defined as $\hat{L}^k \subseteq \hat{L}^j$, $u^k = u^{j}|_{L^k}$ and $\mbox{dim}(L^k) \leq \mbox{dim}(L^j)$.\footnote{$\mbox{dim}(\cdot)$ yields the dimension of a subspace of ${\mathcal L}$.} It follows from this definition that an identity morphism $\rho_{kk}: s^k \rightarrow s^k$ trivially exists for every object $s^k$. Furthermore, given two morphisms $\rho_{kj}: s^k \rightarrow s^j$ and $\rho_{jl}: s^j \rightarrow s^l$ there exists their composition $\rho_{jl}\circ \rho_{kl} = \rho_{kl}$, since $\hat{L}^k \subseteq \hat{L}^j \subseteq \hat{L}^l$, $ \mbox{dim}(L^k) \leq  \mbox{dim}(L^j) \leq  \mbox{dim}(L^l)$ and by transitivity of the restrictions $u^k = u^{j}|_{L^k}$ and $u^j = u^{l}|_{L^j}$ we have that $u^k = u^{l}|_{L^k}$.
\end{itemize}
\end{defi}
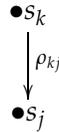
\begin{figure}[h]
\begin{displaymath}
\xymatrix{
 \bullet s_k\ar[d]^{\rho_{kj}}\\
 \bullet s_j}
\end{displaymath}
\caption{Morphism $\rho_{kj}$ from $s^k$ to $s^j$.}
\end{figure}

\noindent We can also define ${\mathcal P}({\mathcal L})$ as the category in which the objects are subsets of ${\mathcal L}$ and a morphism between two objects $f_{AB}: A \rightarrow B$ is defined as $A \subseteq B$.\\

\begin{figure}[h!]
\begin{displaymath}
\xymatrix{
 \bullet A\ar@{^{(}->}[d]\\
 \bullet B}
\end{displaymath}
\caption{Inclusion morphism representing $A \subseteq B$.}
\end{figure}
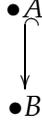

\noindent Let us now define now a functor

\[\Sigma:\mathbf{\mathcal{PR}} \longrightarrow{\mathcal P}({\mathcal L})\]

\noindent which assigns to a problem $s^k = \langle \hat{L}^k, u^k, \hat{{\mathbf X}}^k \rangle$ the subset $\Sigma(s^k)$ of $\mathcal {L}$  defined by

\[\Sigma(s^k)=\{y\in\mathcal{L}\ |\ \Gamma_k(y) \in  \hat{{\mathbf X}}^k \}\]

\noindent A {\it section} $\sigma_k$ over $s^k$ is simply the assignment of the elements of $\Sigma(s^k)$ to $s^k$:

\[\sigma_k:s^k \mapsto \Sigma(s^k).\]

\noindent Given two problems, $s^k = \langle \hat{L}^k, u^k, \hat{{\mathbf X}}^k \rangle$ and  $s^{j} = \langle \hat{L}^{j}, u^{j}, \hat{{\mathbf X}}^{j} \rangle$, let us write $s^k \triangleleft s^{j}$ iff  there exists a morphism $\rho$ in $\mathbf{\mathcal{PR}}$, $\rho: s^k \rightarrow s^{j}$. That is, $s^k$ is a restriction of $s^{j}$.\\

\noindent Let us define $r^{j}_{k}: \Sigma(s^{j}) \rightarrow \Sigma(s^k)$ such that to $\Sigma(s^{j})$ it assigns $\Sigma(s^k)$. Given a section over $s^{j}$, $r^{j}_{k}$ yields a section corresponding to its sub-problem $s^k$.\\

\noindent The following proposition then shows that the functor $\Sigma$ possesses an important property that will be crucial for formalizing the possibility of patching up local problems and yielding a ``larger'' one:

\begin{propi}~\label{sheaf0}
$\Sigma$ is a presheaf.
\end{propi}
\noindent {\bf Proof}: {\it $\Sigma:\mathbf{\mathcal{PR}} \rightarrow  {\mathcal P}({\mathcal L})$ is a functor. We can analyze its behavior by means of $r^{j}_{k}$:}
\begin{itemize}
\item {\it For any $s^k \in \mbox{Obj}(\mathbf{\mathcal{PR}})$, since $s^k \triangleleft s^k$, $r^{k}_{k}= \mbox{Id}_{\Sigma(s^k)}$. }
\item {\it If $s^k \triangleleft s^{j} \triangleleft s^{l}$ then $s^k \triangleleft s^{l}$. Thus, $r^{j}_{k} \circ r^{l}_{j}$$=$$r^{l}_{k}$.}
\end{itemize}
{\it This means that $\Sigma: \mathbf{\mathcal{PR}} \rightarrow {\mathcal P}({\mathcal L})$ is a {\em contravariant} functor. Or, in categorical terms, a} {\em presheaf}. \ $\Box$\\

\noindent Consider now a family $\{s^k = \langle \hat{L}^{k}, u^{k}, \hat{{\mathbf X}}^{k}\rangle \}_{k \in K} \subseteq \mbox{Obj}(\mathbf{\mathcal{PR}})$. It is said to be a {\em cover} of an object $s^j = \langle \hat{L}^j, u^j, \hat{{\mathbf X}}^j\rangle$ of $\mbox{Obj}(\mathbf{\mathcal{PR}})$ if $s^k \triangleleft s^j$ for each $k \in K$ and $\hat{L}^j \subseteq \cup_{k \in K} \hat{L}^{k}$. That is, a problem $s^j$ gets covered by the family $\{s^k\}_{k \in K}$ if the domain of problem $s^j$ is included in the union of the domains of the problems of the family and furthermore, each $s^k$ is a restriction of $s^j$.\\

\noindent The family of sections $\{\sigma_k\}_{k \in K}$ is said to be {\em compatible} if for any pair $k, l \in K$, given $\Sigma(s^k) = X^k$ and $\Sigma(s^l) = X^l$,

$$ \Gamma_k(X^k) \cap \Gamma_l(X^k) = \Gamma_k(X^l) \cap \Gamma_l(X^l)$$

\noindent Given a cover $\{s^k\}_{k \in K}$ of a problem $s^j$ with compatible sections, $\Sigma$ is then a $K$-{\em sheaf} if there exists a unique $\sigma_j = \Sigma(s^j)$ such that for each $k \in K$,

$$ \sigma_k = \sigma_{j} \cap \Gamma_{k}^{-1}(\hat{L}^k)$$

\noindent That is, intuitively, $\Sigma$ is a K-sheaf if $\sigma_j$  in fact ``glues'' together all the assignments $\sigma_k$ in ${\mathcal P}({\mathcal L})$ within the more general framework of their compatibility. Finally, then, if $\Sigma$ is a K-sheaf for every $\{\sigma_k\}_{k \in K} \subseteq \mbox{Obj}(\mathbf{\mathcal{PR}})$ it is called a {\em sheaf}. \\

\begin{example}
\noindent {\it Let ${\mathcal L}$ to be $\mathbb{R}^3$ (the three-dimensional real Euclidean space) and the utility function:}

$$ U(x,y,z) = 3 - 2x^2 - y^2 - 3z^2$$

\noindent {\it to be maximized over ${\mathcal L}$. This yields a single global solution} $\hat{{\mathbf X}}=\{(0,0,0)\}$.\\

\noindent {\it Now consider two possible local problems:}
\begin{itemize}
\item $L^{1} = \{(x,y,z): z=0\}$, {\it with $u^1(x,y,z)$$=$$ U_{|L^1} =$$3- 2 x^2 - y^2$ to be maximized over $ \hat{L}^{1}= \{(x,y,0) \in L^1: x^2 + y^2 =1\}$, the unit circumference in $L^1$. The class of solutions for this problem is $\hat{{\mathbf X}}^{1}= \{(0,1,0), (0, -1,0)\}$.}
\item $L^{2} = \{(x,y,z): (x,y,z) \cdot (1,-1,1)= 0\}$ {\it (i.e. the linear subspace with normal vector $(1,-1,1)$), with $u^2(x,y,z) = 3- 3 x^2 - 4z^2 - 2xz$, the restriction of $U$ on $L^{2}$, to be maximized over $ \hat{L}^{2}= \{(x,y,z): 2x^2 + 2z^2 + 2xz =1\}$, the intersection of the surface of the unit sphere in $\mathbb{R}^3$ with $L^2$. Here the solution set is: $\hat{{\mathbf X}}^{2}= \{(-\sqrt{\frac{1}{3}}, -\frac{1}{2 \sqrt{3}} - \frac{1}{2}, \frac{1}{2 \sqrt{3}} - \frac{1}{2}), (\sqrt{\frac{1}{3}}, \frac{1}{2 \sqrt{3}} + \frac{1}{2}, \frac{1}{2} - \frac{1}{2 \sqrt{3}} )\}$.}
\end{itemize}

\noindent {\it It is easy to see that each solution of problem $1$ minimizes the distance to the projection of the single global solution $(0,0,0)$ on $L^1$. More precisely $\Gamma_1(0,0,0) = \hat{{\mathbf X}}^{1}$. The same is true for problem $2$, since all points in $L^{2}$ are at a Euclidean distance $1$ from the global solution. So, in particular, the elements in $\hat{{\mathbf X}}^{2}$ minimize the distance to the projection of $(0,0,0)$ on $L^{2}$ and thus, $\Gamma_2 (0,0,0) = \hat{{\mathbf X}}^{2}$ }.\\

\noindent {\it Given problems 1 and 2, denoted  $s^i = \langle \hat{L}^i, u^i, \hat{{\mathbf X}}^i \rangle$ for $i=1,2$, we add a new problem $s^{0}$, which is the optimization of $U$ over the surface of the three-dimensional sphere $\hat{L}^{0} = \{(x,y,z): x^2 + y^2 + z^2 = 1\}$ and thus, $\hat{{\mathbf X}}^0 = \{(0,1,0), (0,-1,0)\}$. Suppose that these are the only objects in $\mathbf{\mathcal{PR}}$. We define $\Sigma: \mathbf{\mathcal{PR}} \rightarrow {\mathcal P}({\mathcal L})$, summarized by the following table (each row being a section $\sigma_i$, $i=0,1,2$):}

\begin{center}
\begin{tabular}{|c|c|c|c|c|}
\hline\hline
\mbox{Problems}   & $a_1$ & $b_1$ & $a_2$ & $b_2$ \\ \hline
$s^1$             & $X$   & $-$   & $X$   & $-$ \\ \hline
$s^2$             & $-$   & $X$   & $-$   & $X$ \\ \hline
$s^0$             & $X$   & $-$   & $X$   & $-$ \\ \hline

\end{tabular}
\end{center}

\noindent {\it The range of $\Sigma$ is based only of four elements in} ${\mathcal L}$:

$$ a_1 = (0, 1,0) \ \ \ a_2 = (0, -1, 0)$$

\noindent {\it and}

$$ b_1 = (-\sqrt{\frac{1}{3}}, -\frac{1}{2 \sqrt{3}} - \frac{1}{2}, \frac{1}{2 \sqrt{3}} - \frac{1}{2})\ \ \ b_2 = (\sqrt{\frac{1}{3}}, \frac{1}{2 \sqrt{3}} + \frac{1}{2}, \frac{1}{2} - \frac{1}{2 \sqrt{3}} )$$

\noindent {\it where $a_1$ and $a_2$ are the $\mathbb{R}^3$ solutions of problems $s^0$ and $s^1$ while $b_1$ and $b_2$ are those of $s^2$.}\\

\noindent {\it It is easy to check that $s^i \triangleleft s^{0}$ for $i=1,2$, since on one hand each problem $s^i$ can be seen as the maximization of $U$ restricted to subsets of the domain of problem $s^{0}$. On the other hand, $\Sigma(s^{0})$ restricted to each $s^i$ yields $\Sigma(s^i)$. In fact, for $s^1$ it is clear that this is the case. For $s^2$, let us note that $b_1, b_2$ are the solutions of the problem $s^0$ restricted to $\hat{L}^2$, seen as the inverse projection over the surface $\hat{L}^0$.\\

\noindent Furthermore, $\{\sigma_1, \sigma_2\}$ is a compatible family of sections. Notice that $\hat{L}^1 \cap \hat{L}^2$ does not include the solutions to either problem. But then the projections of $\hat{\mathbf{X}}^1$ and $\hat{\mathbf{X}}^2$ on $\hat{L}^1 \cap \hat{L}^2$ are both $\emptyset$, and thus the sections satisfy, trivially, the compatibility condition. This means that $\Sigma$ satisfies the sheaf condition.}

\end{example}

\noindent Summarizing the discussion up to this point, we can say that given a category of problems $\mathbf{\mathcal{PR}}$ over a space ${\mathcal L}$, it is typically desirable to be able to obtain a sheaf $\Sigma: \mathbf{\mathcal{PR}} \rightarrow {\mathcal P}({\mathcal L})$, such that for any problem $\mathbf{s}^j$, covered by any compatible family of sub-problems, $\{s^k\}_{k \in K}$, $\Sigma(\mathbf{s}^j) \cap \Gamma_k^{-1}(\hat{L}^k) = \Sigma(s^k)$ for $k \in K$. 

\section{A Categorical Representation of Games}

\noindent Let us now consider, instead of the coordination of different local decision problems, the coordination of games. That is, decision problems involving several agents, instead of a single one. Thus, the approach discussed in this section generalized the sheaf-theoretical framework presented above.\\

\noindent Let us consider a category $\mathcal{G}$ of {\em games}. Each object $G$ in the category  corresponds to a game $G = \langle (I_G, S_G, \mathbf{O}_G, \rho_G),\pi_G \rangle$, where\\
\begin{itemize}
\item $(I_G, S_G, \mathbf{O}_G, \rho_G)$ is a game form:
\begin{itemize}
\item $I_G$ is the class of players. 
\item $S_G = \prod_{i \in I_G} S_i^G$ is  the {\em strategy set} of the game, where $S_i^G \subseteq S_i$ is the set of strategies that player $i$ can deploy in game $G$, for each $i \in I_G$.\footnote{$S_i$ is the set of all the strategies that player $i$ can play in the games in which she participates.}
\item $\mathbf{O}_G$ is the class of {\em outcomes} of the game and $\rho_G: S_G \rightarrow \mathbf{O}_G$ is a one-to-one function that associates each profile of strategies in the game with one of its outcomes.
\end{itemize}
\item $\pi_G = \prod_{i \in I} \pi_i^G$ is a {\em profile of payoff functions}, where $\pi_i^G: \mathbf{O}_G \rightarrow \mathbb{R}^+$ is the payoff function of player $i$ in game $G$, for each $i \in I_G$.

\end{itemize}
\noindent A game is defined in terms of the interactions of {\em players}. Each player can be seen as described in terms of the strategies she can play and the payoffs she can receive from the results of her action (jointly with those of the other players).\\

\noindent We can define a category $\mathcal{G}$, where the objects are games. Given two games 
$$G = \langle (I_G, S_G, \mathbf{O}_G, \rho_G),\pi_G \rangle \ \ \ \ \mbox{and} \ \ \ \ \ G^{\prime} = \langle (I_{G^{\prime}}, S_{G^{\prime}}, \mathbf{O}_{G^{\prime}}, \rho_{G^{\prime}}),\pi_{G^{\prime}} \rangle,$$
\noindent a morphism of games 
$$ G \rightarrow G^{\prime}$$

\noindent is such that:

\begin{itemize}
\item $I_G \subseteq I_{G^{\prime}}$.
\item $S_i^G \subseteq S_i^{G^{\prime}}$ for each $i \in I_G$.
\item There exist two functions, an inclusion $p^{\mathbf{O}_{G^{\prime}}}_{\mathbf{O}_G}: SO_{G^{\prime}} \hookrightarrow \mathbf{O}_G$ for $SO_{G^{\prime}} \subseteq \mathbf{O}_{G^{\prime}}$ and a projection $p^{S_{G^{\prime}}}_{S_G}: S_{G^{\prime}} \rightarrow S_G$, i.e. $p^{S_{G^{\prime}}}_{S_G}(s_1^{G^{\prime}}, \ldots, s_i^{G^{\prime}}, \ldots, s_{|I_{G^{\prime}}|}) \in \prod_{i \in I_G} S_i^{G^{\prime}} = S_{G}$. These functions verify the following condition:
\begin{itemize}
\item For every $s^{\prime} \in S_{G^{\prime}}$, $s = p^{S_{G^{\prime}}}_{S_G}(s^{\prime}) \in S_G$ is such that $\rho_{G}(s) =  p^{\mathbf{O}_{G^{\prime}}}_{\mathbf{O}_G}(\rho_{G^{\prime}}(s^{\prime}))$.
\end{itemize}
\end{itemize}
\noindent Thus, if a morphism $G \rightarrow G^{\prime}$ exists, $G$ can be conceived as a {\em subgame form} of $G^{\prime}$.\\

\noindent To complete the characterization of $\mathcal{G}$ notice that it is immediate that we can define {\em pushouts} and an {\em initial object} in this category:\\

\begin{itemize}
\item {\bf Pushouts}: Consider three objects $G$, $G^{\prime}$ and $G^{\prime \prime}$ and morphisms $G \stackrel{f}{\rightarrow} G^{\prime}$ and $G \stackrel{g}{\rightarrow} G^{\prime \prime}$. Then, take the coproduct of $G^{\prime}$ and $G^{\prime \prime}$, denoted $G^{\prime} + G^{\prime \prime}$, obtained as the direct sums of the strategies sets and the outcomes of both games. By identifying the subgame forms of $G^{\prime}$ and $G^{\prime \prime}$ corresponding to $G$ we obtain the {\em pushout} of 
$$G^{\prime}  \ \ \stackrel{f}{\leftarrow} \ \ G \ \ \stackrel{g}{\rightarrow} \ \ G^{\prime \prime}$$

\item {\bf Initial object}: Consider the {\em empty game} $G^{\emptyset}$, where $I_{G^{\emptyset}} = \emptyset$ and consequently  $S_{G^{\emptyset}} = \emptyset$ and $\mathbf{O}_G = \emptyset$ (thus $\pi_{G^{\emptyset}}$ must be the empty function). It is immediate to see that $G^{\emptyset} \rightarrow G$ for every $G$ in $\mathcal{G}$.
\end{itemize}
\noindent Then we have 
\begin{proposition}
$\mathcal{G}$ is a category with {\em colimits}.
\end{proposition}
\noindent Since $\mathcal{G}$ is a category with colimits we can define {\em cospans} in it. Consider again three objects $G$, $G^{\prime}$ and $G^{\prime \prime}$ and two morphisms $G  \stackrel{f}{\rightarrow}  G^{\prime \prime} \stackrel{g}{\leftarrow} G^{\prime}$. This is called a cospan from $G$ to $G^{\prime}$. The interpretation of such a cospan is that $G$ and $G^{\prime}$ are subgame forms of the same game ($G^{\prime \prime}$).\\


\noindent We can conceive each game $G$ in $\mathcal{G}$ as a {\em box}, $G = (\mbox{in}^G, \mbox{out}^G)$, where $\mbox{in}^G$ and $\mbox{out}^G$ are, respectively {\em input} and {\em output} ports. $\mbox{in}^G$ has type $\mathbf{O}_G$, i.e. the input is an outcome of $G$. In turn, the $\mbox{out}^G$ port has type $S_G$, being each output a profile in $G$.\\

\noindent Notice that each player $i$ can be conceived as a game $(\mbox{in}^i, \mbox{out}^i)$, where $\mbox{in}^i$ has type $\cup_{G: i \in I_G} \mathbf{O}_G$ and $\mbox{out}^i$ has type $S_i$.\\ 

\noindent Up to this point, our definition of morphisms in $\mathcal{G}$ does not involve the payoffs. They can be incorporated by redefining the games as {\em modal boxes}, in which an additional component are the {\em internal states} of the game. More precisely, given any $G$ and the class of its internal states, $\Sigma_G$, we can identify $G$ as a triple $\langle \mbox{in}^G, \mbox{out}^G, \Sigma_G \rangle$, associated to two correspondences:

\begin{itemize}
\item {\bf payoff}: $\phi_G^1: \bar{\mbox{in}}^G \times \Sigma_G \rightarrow \mathbb{R}^{+^{\mathbf{O}_G}}$, such that for the vector $o \in \bar{\mbox{in}}^G$ (the vector of all possible inputs of $G$, each entry being an outcome of the game) and state $\sigma$, $\phi_G^1(o,\sigma) =  (\pi_G^i(o))_{o \in \mathbf{O}_G}$. That is, it yields the vector of payoffs corresponding to all the outcomes of $G$.
\item {\bf choice}: $\phi_G^2: \Sigma_G \rightarrow \bar{\mbox{out}}^G$, such that for any state $\sigma$, $\phi_G^2(\sigma) = s \in \bar{\mbox{out}}^G$ (the class of all possible strategy profiles in $S_G$) is a profile of strategies that may be chosen at that state.  
\end{itemize}
 
\noindent Particularly relevant for our analysis is the definition of the internal states of each player $i$, $\Sigma_i$. Consider a game $G$ such that $i \in I_G$, and a sequence of morphisms in $\mathcal{G}$ 
$$ G_{i}^{0} \rightarrow G_i^{1} \rightarrow \ \ \ \ldots \ \ \ \rightarrow G_i^{n-1} \rightarrow G_i^n$$
\noindent where $G_{i}^0$ is a game in which $i$ is the only player and $G = G_i^n$. We identify the state of player $i$ when playing $G$ as a sequence $\sigma^i_G =$$\langle \sigma^i_0, \ldots, \sigma^i_{n-1}\rangle$, where $\sigma^i_k \in \Sigma_{G_{i}^k}$, for $k= 0 \ldots, n-1$. Then, a distinguished object $\sigma^i_* \in \Sigma_i$ is defined, such that $\sigma^i_G$ is one of its initial segments.\footnote{Thus, $\sigma^i_*$ has a {\em forest} structure.} \\

\noindent Therefore, for each game $G$, $\sigma^i_*$ can be instantiated yielding the corresponding state, and therefore the payoffs and the choices of player $i$ in the game. The state $\sigma_G$ of the entire game just obtains as the profile of states of its players.\\

\noindent A simple example is $\sigma^i_{G^n}$ yielding as payoff for $i$ the product of the payoffs she gets in the subgames of $G^n$. This case will be elaborated a bit more in Example 1, below.\\


\noindent We can define the category of cospans in $\mathcal{G}$, denoted $\mbox{cospan}_{\mathcal{G}}$ which has a symmetric monoidal structure. Its objects are the same as those of $\mathcal{G}$ and a morphism $G \stackrel{h}{\rightarrow} G^{'}$ is a cospan from $G$ to $G^{\prime}$, indicating that there exists a game of which $G$ and $G^{\prime}$ are subgame forms. Thus, morphisms in $\mbox{cospan}_{\mathcal{G}}$ are actually isomorphisms.\\

\noindent Given two morphisms in $\mbox{cospan}_{\mathcal{G}}$, $G  \stackrel{f}{\rightarrow}  G^{\prime}$ and $G^{\prime}  \stackrel{g}{\rightarrow} G^{\prime \prime}$ there exists a morphism $G  \stackrel{g\circ f}{\rightarrow}  G^{\prime \prime}$ that obtains as a composition of the corresponding cospans.

\noindent The monoidal structure of $\mbox{cospan}_{\mathcal{G}}$ is given by:

\begin{itemize}
\item The unit is $G^{\emptyset}$, the initial object in $\mathcal{G}$.

\item The monoidal product of $G$ and $G^{\prime}$, is the coproduct $G + G^{\prime}$.
 
 \end{itemize}

\noindent We now present a diagram language for open games. We start by considering the symmetric monoidal category $\mathbf{W}_{\mathcal{G}}$. By definition, we have that:

$$ \mathbf{W}_{\mathcal{G}} \ = \ \mbox{cospan}_{\mathcal{G}}$$

\noindent Each object, i.e. a game $G$, is seen as a $\langle \mbox{in}^G, \mbox{out}^G, \Sigma_G \rangle$-labeled {\em interface}, satisfying $\phi_G^1$ and $\phi_G^2$. On the other hand, morphisms $G \rightarrow C \leftarrow G^{\prime}$, are called $\langle \mbox{in}, \mbox{out}, \Sigma \rangle$-labeled {\em wiring diagrams}. The interpretation is that $C$ is the overarching game that connects the subgames (not just the game forms) $G$ and $G^{\prime}$.\\

\noindent We write $\psi: G_1, G_2,\ldots, G_n \rightarrow \bar{G}$ to denote the wiring diagram $\phi: G_1 + G_2 + \ldots + G_n \rightarrow \bar{G}$. We can, in turn see this as
$$ G_1 + G_2 + \ldots + G_n \stackrel{f}{\rightarrow} C \stackrel{\bar{f}}{\leftarrow} \bar{G}$$
\newline
\noindent which indicates that, being $f$ and $\bar{f}$ isomorphisms,

\begin{proposition}
$\bar{G}$ is the minimal game that includes the direct sum of $G_1, \ldots, G_n$ as a subgame.
\end{proposition}

\section{Hypergraph Categories and Equilibria}

\noindent We define a {\em hypergraph category} $\langle \mathcal{G}, \mathbf{\mbox{Eq}} \rangle$ with $\mathbf{\mbox{Eq}}: \mathbf{W}_{\mathcal{G}} \rightarrow \prod_i S_i$, such that, for every object $G$ in $\mathbf{W}_{\mathcal{G}}$, $\mathbf{\mbox{Eq}}(G)$ is a class of vectors in $\prod_{i \in I} S_i^G$,  the strategy set of game $G$. We assume that $\mathbf{\mbox{Eq}}(G)$ is a class of {\em equilibria} of $G$, for some notion of equilibrium (as for instance, dominant strategies equilibrium, admissible strategies, or Nash equilibrium).\\

\begin{example}

\noindent Consider two games, $G$ between players $1$ and $2$:\footnote{This a {\em Battle of the Sexes} game, where $S_1 = S_2 = \{\mbox{Bx}, \mbox{Bll}\}$.}\\

\begin{game}{2}{2}
              & $\mbox{Bx}$ & $\mbox{Bll}$\\
$\mbox{Bx}$   & \colorbox{red}{$2,1$}       & $0,0$\\
$\mbox{Bll}$   & $0,0$       & \colorbox{red}{$1,2$}\\
\end{game}
\newline

\noindent and $G^{\prime}$ between players $2$ and $3$:\footnote{A {\em Prisoner's Dilemma}, where $S_2 = S_3 = \{C, D\}$.}\\

\begin{game}{2}{2}
              & $\mbox{C}$ & $\mbox{D}$\\
$\mbox{C}$   & $2,2$       & $0,3$\\
$\mbox{D}$   & $3,0$       & \colorbox{red}{$1,1$}\\
\end{game}
\newline

\noindent The corresponding wiring diagram is:\\

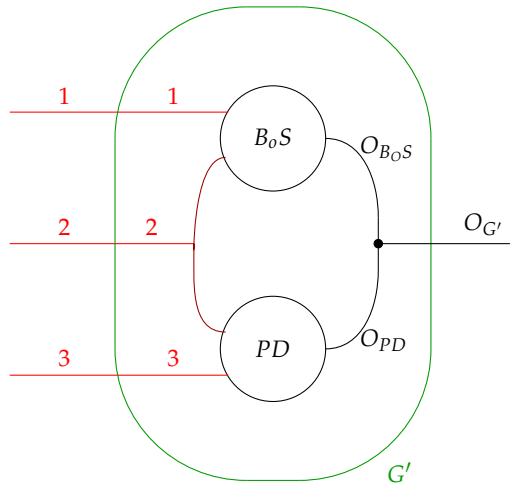
\begin{figure}[h!]
	\centering
	\footnotesize
	
	\begin{tikzpicture}[scale=0.7]
	
	\draw (5,5) circle (1cm);
	\node at (5,5) {$B_oS$};
	\draw (5,1) circle (1cm);
	\node at (5,1) {$PD$};
	
	\draw[red] (0,5.5) -- (4.15,5.5);
	\draw[red] (0,3.0) -- (3.50,3.0);
	\draw[red] (0,0.5) -- (4.15,0.5);
	
	\node[red,anchor=south] at (1.0375,5.5) {$1$};
	\node[red,anchor=south] at (3.1125,5.5) {$1$};
	
	\node[red,anchor=south] at (1.0375,3.0) {$2$};
	\node[red,anchor=south] at (2.7,3.0) {$2$};
	
	
	\node[red,anchor=south] at (1.0375,0.5) {$3$};
	\node[red,anchor=south] at (3.1125,0.5) {$3$};
	
		\draw [red!60!black] (3.50,3.0) to[out=270,in=180] (4.10,4.64);
		\draw [red!60!black] (3.50,3.0) to[out=270,in=180] (4.10,1.32);
	
	\draw [green!60!black] (2,5) to[out=90,in=180] (4.5,7.5);
	\draw [green!60!black] (5.5,7.5) to[out=0,in=90] (8,5);
	\draw [green!60!black] (4.5,7.5) -- (5.5,7.5);
	
	\draw [green!60!black] (2,1) to[out=270,in=180] (4.5,-1.5);
	\draw [green!60!black] (5.5,-1.5) to[out=0,in=270] (8,1);
	\draw [green!60!black] (4.5,-1.5) -- (5.5,-1.5);
	
	\draw [green!60!black] (2,1) -- (2,5);
	\draw [green!60!black] (8,1) -- (8,5);
	
	\node[green!60!black,anchor=north west] at (7,-1) {$G'$};
	
	\draw [] (6,5) to[out=0,in=90] (7,3.5);
	\draw [] (6,1) to[out=0,in=270] (7,2.5);
	
	\draw [] (7,3.5) -- (7,2.5);
	
	\node[anchor=north west] at (6.5,5.2) {$O_{B_OS}$};
	\node[anchor=south west] at (6.5,0.8) {$O_{PD}$};
	
	\filldraw[black] (7,3) circle (0.08cm);
	
	\draw [] (7,3) -- (9.5,3);
	\node[anchor=south] at (9,3) {$O_{G'}$};
	
	\end{tikzpicture}
\end{figure}

\noindent In red we have highlighted $\mathbf{\mbox{Eq}}(G) = \{ (\mbox{Bx,Bx}), (\mbox{Bll, Bll})\}$ and $\mathbf{\mbox{Eq}}(G^{\prime}) = \{ (\mbox{D,D})\}$, where $\mathbf{\mbox{Eq}}$ corresponds to Nash equilibrium.\footnote{Notice that here player $2$, participates in two games.}\\

\noindent Let us represent now $G + G^{\prime}$. We start by building its corresponding game form. We obtain two tables, where the first one corresponds to player $3$ choosing $\mbox{C}$:\\
\begin{game}{2}{4}
              & $\mbox{Bx-C}$ & $\mbox{Bx-D}$ & $\mbox{Bll-C}$ & $\mbox{Bll-D}$ \\
$\mbox{Bx}$   & $\mbox{o}_{1,1}$       & $\mbox{o}_{1,2}$  & $\mbox{o}_{1,3}$    & $\mbox{o}_{1,4}$ \\
$\mbox{Bll}$   & $\mbox{o}_{2,1}$       & $\mbox{o}_{2,2}$ &  $\mbox{o}_{2,3}$   & $\mbox{o}_{2,4}$
\end{game}
\newline\\


\noindent and another corresponding to player $3$ choosing $\mbox{D}$:\\
\begin{game}{2}{4}
              & $\mbox{Bx-C}$ & $\mbox{Bx-D}$ & $\mbox{Bll-C}$ & $\mbox{Bll-D}$ \\
$\mbox{Bx}$   & $\mbox{o}^{\prime}_{1,1}$       & $\mbox{o}^{\prime}_{1,2}$  & $\mbox{o}^{\prime}_{1,3}$    & $\mbox{o}^{\prime}_{1,4}$ \\
$\mbox{Bll}$   & $\mbox{o}^{\prime}_{2,1}$       & $\mbox{o}^{\prime}_{2,2}$ &  $\mbox{o}^{\prime}_{2,3}$   & $\mbox{o}^{\prime}_{2,4}$
\end{game}
\newline\\


\noindent For instance, $\mbox{o}_{11}$ indicates that $1$ and $2$ go to Box and $2$ and $3$ Cooperate. On the other hand, $\mbox{o}^{\prime}_{1,1}$ indicates that, again $1$ and $2$ go to Box, but while $2$ keeps Cooperating, $3$ Defects. The other entries can be interpreted likewise.\\

\noindent Suppose that the internal states of the players, $\sigma^1_*, \sigma^2_*$ and $\sigma^3_*$ are such that instantiated on $G + G^{\prime}$ yield the following payoffs and choices:\\
\newline
\noindent If $3$ chooses $\mbox{C}$:\\

\begin{game}{2}{4}
              & $\mbox{Bx-C}$ & $\mbox{Bx-D}$ & $\mbox{Bll-C}$ & $\mbox{Bll-D}$ \\
$\mbox{Bx}$    &$2,1 \times 2, 2$           & $2,1 \times 3,0$  & $0, 0 \times 2, 2$   &  $0, 0 \times 3, 0$ \\
$\mbox{Bll}$   & $0,0 \times 2, 2$         & $0, 0 \times 3, 0$ & $1, 2\times 2, 2$ & $1, 2 \times 3, 0$
\end{game}
\newline\\

\noindent while if $3$ chooses $\mbox{D}$:\\

\begin{game}{2}{4}
              & $\mbox{Bx-C}$ & $\mbox{Bx-D}$ & $\mbox{Bll-C}$ & $\mbox{Bll-D}$ \\
$\mbox{Bx}$    &$2,1 \times 0, 3$           & \colorbox{red}{$2,1 \times 1,1$}  & $0, 0 \times 0, 3$   &  $0, 0 \times 1, 1$ \\
$\mbox{Bll}$   & $0,0 \times 0, 3$         & $0, 0 \times 1, 1$ & $1, 2\times 0, 3$ & \colorbox{red}{$1, 2 \times 1, 1$}
\end{game}
\newline\\

\noindent In words, players $1$ and $3$ keep the payoffs they get in the subgames, while $2$ takes the product of the payoffs in $G$ and $G^{\prime}$. In red, we have highlighted the equilibria of $G + G^{\prime}$, under this specification.

\end{example}
\noindent Let us define an operation $\hat{\cup}$ such that given two equilibria $s \in \mathbf{\mbox{Eq}}(G)$ and $s^{\prime} \in \mathbf{\mbox{Eq}}(G^{\prime})$, yields a new profile $s-s^{\prime} \in \mathbf{\mbox{Eq}}(G) \hat{\cup} \mathbf{\mbox{Eq}}(G^{\prime})$ verifying that for each player $i \in I_G \cap I_{G^{\prime}}$, a new strategy obtains combining  $s_i$ and $s_i^{\prime}$, while in on all other cases the individual strategies are the same as in $G$ and $G^{\prime}$. Furthermore, $\pi^{G \hat{\cup} G^{\prime}}_i(s-s^{\prime}) = \pi^{G}_i(s) \times \pi^{G^{\prime}}_i(s^{'})$ for $i \in I_G \cap I_{G^{\prime}}$.\footnote{An alternative yielding also Proposition~\ref{3} obtains if, instead, we take $\pi^{G \hat{\cup} G^{\prime}}_i(s-s^{\prime}) = \pi^{G}_i(s) + \pi^{G^{\prime}}_i(s^{'})$ for $i \in I_G \cap I_{G^{\prime}}$.}\\

\noindent In our example, since $\mathbf{\mbox{Eq}}(G + G^{\prime}) = \{ (\mbox{Bx, Bx-D, D}), (\mbox{Bll, Bll-D, D})\}$, we have that\\
$$\mathbf{\mbox{Eq}}(G) \hat{\cup} \mathbf{\mbox{Eq}}(G^{\prime}) = \mathbf{\mbox{Eq}}(G + G^{\prime}).$$\\

\noindent This example illustrates the following claim:\\

\begin{proposition}\label{3}
For any pair of games $G$ and $G^{\prime}$, $\mathbf{\mbox{Eq}}(G) \hat{\cup} \mathbf{\mbox{Eq}}(G^{\prime}) = \mathbf{\mbox{Eq}}(G + G^{\prime})$.
\end{proposition}
\noindent {\bf Proof}: {\it Trivial. If $I_G \cap I_{G^{\prime}} = \emptyset$, $G + G^{\prime} = G \cup G^{\prime}$ with $G \cap G^{\prime} = \emptyset$. Thus, each equilibrium of $G + G^ {\prime}$ is just the disjoint combination of equilibria in $G$ and $G^{\prime}$.

\noindent If, on the other hand, $I_G \cap I_{G^{\prime}} \neq \emptyset$, given $i \in I_G \cap I_{G^{\prime}}$, her strategy set in $G+ G^{\prime}$ is $S_i^{G} \times S_i^{G^{\prime}}$, where $S_i^G$ and $S_i^{G^{\prime}}$ are her strategy sets in $G$ and $G^{\prime}$, respectively. Now suppose that $s_i^G$ and $s_i^{G^{\prime}}$ are equilibrium strategies of $i$ in the individual games but that $(s_i^G, s_i^{G^{\prime}})$ does not belong to an equilibrium in $G+G^{\prime}$. Then, there exist an alternative combined strategy $(\hat{s}_i^{G}, \hat{s}_i^{G^{\prime}})$ such that on the new profile  $\pi_i$ yields a higher payoff, but since this equilibrium can be decomposed in two profiles, one in $G$ and the other in $G^{\prime}$, the payoff of $i$ is the product of the payoffs over those two profiles. But then either $\hat{s}_i^{G}$ yields a higher payoff than $s_i^G$ or $\hat{s}_i^{G^{\prime}}$ yields a higher payoff than $s_i^{G^{\prime}}$ (recall that they are all positive real numbers). Thus, either $s_i^G$ or $s_i^{G^{\prime}}$ is not an equilibrium in the corresponding game. Absurd.} $\Box$\\

\noindent If we denote $+$ the monoidal operation in $\mathbf{W}_{\mathcal{G}}$, if we take $\otimes = \hat{\cup}$ as monoidal operation in $\prod_i S_i$, Proposition~\ref{3} indicates that there exist a trivial {\em natural isomorphism}

$$\mathbf{\mbox{Eq}}(G) \otimes \mathbf{\mbox{Eq}}(G^{'}) \ \rightarrow \ \mathbf{\mbox{Eq}}(G + G^{'})$$

\noindent Furthermore, taking the unit in $\prod_i S_i$ to be the empty set, we have also that $\emptyset= \mathbf{\mbox{Eq}}(G^{\emptyset})$, where $G^{\emptyset}$ is the initial object in $\mathcal{G}$ and thus in $\mathbf{W}_{\mathcal{G}}$.\\

\noindent We have that

\begin{proposition}\label{4}
$\mathbf{\mbox{Eq}}$ is a lax monoidal functor.
\end{proposition}

\noindent Thus, the corresponding algebra allows to associate the composition of games with the equilibria of the components.\\


\noindent Proposition~\ref{4} depends critically on the possibility of defining $\otimes$ in terms of a function $\mathbf{f}$, defined as follows. Given a player $i \in I_G \cap I_{G^{\prime}}$, a combined strategy $s_i - s^{\prime}_i$ is such that for $s = (s_i, s_{-i}) \in \mbox{Eq}(G)$ and $s^{\prime} = (s^{\prime}_i, s^{\prime}_{-i}) \in \mbox{Eq}(G^{\prime})$, satisfying $\pi_i(s - s^{\prime}) = \mathbf{f}(\pi^G_i(s), \pi^{G^{\prime}}_i(s^{\prime}))$ and with $s - s^{\prime} \in \mbox{Eq}(G + G^{\prime})$. As we saw above if $\mathbf{f}$ is the arithmetic product or sum, $\mbox{Eq}$ will be indeed a lax monoidal functor.\\

\noindent But this restricts the compositionality of games to just trivial cases.  We are interested in more general and non-obvious cases. In order to do that consider an alternative characterization of the hypergraph category $\langle \mathcal{G},\mathbf{\mbox{Eq}}\rangle$:\\
$$ \mathbf{\mbox{Eq}}: \mathbf{\mbox{W}}_{\mathcal{G}} \rightarrow \prod_i S_i \times \cup_{G \in \mathbf{\mbox{Obj}}(\mathcal{G})} \Sigma_G$$
\newline
\noindent Furthermore, we need another definition of $\otimes$:\\
$$ \otimes: (\prod_i S_i \times \cup_{G \in \mathbf{\mbox{Obj}}(\mathcal{G})} \Sigma_G ) \ \times \ (\prod_i S_i \times \cup_{G \in \mathbf{\mbox{Obj}}(\mathcal{G})} \Sigma_G ) \ \ \rightarrow \ \ \prod_i S_i \times \bigcup_{G \in \mathbf{\mbox{Obj}}(\mathcal{G})} \Sigma_G$$
\newline
\noindent such that given two games $G$ and $G^{\prime}$ with $s \in \prod_{i \in I_G} S_i$ and $\sigma_G$, and $s^{\prime} \in \prod_{i \in I_{G^{\prime}}} S_i$ and $\sigma_{G^{\prime}}$ we have:\\
$$(s, \sigma_G) \otimes (s^{\prime}, \sigma_{G^{\prime}}) = (\bar{s}, \sigma_{G + G^{\prime}}) \in \prod_{i \in I_{G + G^{\prime}}} S_i \times \Sigma_{G + G^{\prime}}$$
\newline
\noindent where $\bar{s} \in S_{G + G^{\prime}}$ is a Nash equilibrium if and only if $s$ and $s^{\prime}$ are Nash equilibria of $G$ and $G^{\prime}$ respectively.\\

\noindent $\otimes$ is well-defined. To see this, just recall that, by definition $G + G^{\prime}$ obtains in terms of the game forms of $G$ and $G^{\prime}$ (the strategy sets and the outcomes), allowing different possible internal states and thus payoffs. The view of games as boxes presented in Section 4 indicates that there exist sequences of internal states of games, in parallel to sequences of morphisms between games, allowing to define $\sigma_{G + G^{\prime}}$, and thus payoffs that make $\bar{s}$ a Nash equilibrium if $s$ and $s^{\prime}$ are also equilibria.\\ 

\noindent We can see that $\prod_i S_i \times \bigcup_{G \in \mathbf{\mbox{Obj}}(\mathcal{G})} \Sigma_G$ with $\otimes$, defined as above can be seen as a monoidal category, with morphisms defined in terms of those of $\mathcal{G}$, with $(\emptyset, \emptyset)$ as its initial object. It allows to define $\mathbf{\mbox{Eq}}$ in such a way that by definition:\\

\begin{proposition}~\label{5}
$\mathbf{\mbox{Eq}}$ is a lax functor satisfying $ \mathbf{\mbox{Eq}}(G + G^{\prime}) \ = \ \mathbf{\mbox{Eq}}(G) \otimes \mathbf{\mbox{Eq}}(G^{\prime})$.
\end{proposition} 

\section{A more general model}

\noindent $\langle \mathcal{G}, \mbox{Eq}\rangle$, in any of the two versions of $\mbox{Eq}$ seems too rigid to capture the dynamics of economic interactions. A more flexible structure is needed.\\

\noindent Let us start with the category of {\em polynomial functors}, $\mathbf{Poly}$:

\begin{itemize}
\item Its objects have the following general form:
$$p = \sum_{i \in I} y^{p[i]}$$
\noindent where each term $y^{p[i]}$ is a functor with domain $p[i]$ into $\mathbf{Set}$. Each $i$ can be conceived as a {\em problem} while $p[i]$ is a set of its {\em solutions}.
\item Given $p = \sum_{i \in I} y^{p[i]}$ and $q = \sum_{j \in J} y^{q[j]}$ a morphism $\phi: p \rightarrow q$ is $\phi = (\phi_{\rightarrow}, \phi^{\leftarrow})$ such that 
\begin{itemize}
\item $\phi^{\rightarrow}: I \rightarrow J$ and,
\item $\phi^{\leftarrow}: q[\phi^{\rightarrow}(i)] \mapsto p[i]$.
\end{itemize}
\noindent That is, $\phi$ sends problems of $I$ into problems of $J$ and then the corresponding solutions in $q$ back to the solutions in $p$.
\end{itemize}

\noindent We can conceive any $p \in Ob(\mathbf{Poly})$ as an {\em interface} between inputs and outputs, being the inputs problems and the outputs their solutions. There are different ways of creating new interfaces up from other interfaces. We focus on the following construction:

\begin{itemize}
\item $[p, q] = \sum_{\phi: p \rightarrow q} y^{\sum_{i \in I} q[\phi^{\rightarrow}(i)]}$, an {\it internal hom} in $\mathbf{Poly}$. It can be seen as a process
that takes as inputs ({\em problems}) the morphisms from $p$ to $q$ and as outputs ({\em solutions}) all the possible solutions to the images of $p$ in $q$.
\item  Given $[p,q]$, a $[p, q]-\mathbf{Coalg}$ is a category in which each object is triple $\langle s, \rho, \mu \rangle$:
\begin{itemize}
\item $s \in S$, where $S$ is a space of {\em states}, capturing the dynamics of the interface,
\item $\rho: s \mapsto (\phi, i, q[\phi^{\rightarrow}(i)])$. That is, it assigns to the current state one of the solutions in $[p,q]$,
\item $\mu$ updates the state in response to that pattern, i.e. $\mu(\phi, i, q[\phi^{\rightarrow}(i)]) = s^{\prime} \in S$.
\end{itemize}
\end{itemize}

\noindent Consider now a category $\mathbf{\mathbb{O}rg}$ defined as follows: 
\begin{itemize}
\item $Ob(\mathbf{\mathbb{O}rg})= Ob(\mathbf{Poly})$ and, \item $Morph(\mathbf{\mathbb{O}rg})= [p, q]-\mathbf{Coalg}$.
\end{itemize}

\noindent This means that two interfaces (connecting problems with their solutions) $p$ and $q$ are related by dynamic procedures of reconnection between them.\\

\noindent Our generalized model, covering {\em both} $\mathcal{PR}$ and $\langle \mathcal{G}, \mbox{Eq}\rangle$ is a category $\mathbf{\mathcal{PR - G}}$ based on $\mathbf{\mathbb{O}rg}$ such that, briefly:
\begin{itemize}
\item for each object $a$ it corresponds $p_a$ in $\mathbf{\mathbb{O}rg}$,
\item for objects $a_1, \ldots , a_n, b$ there corresponds a
$[p_{a_1} \otimes \ldots \otimes p_{a_n}, p_b]-\mathbf{Coalg}$ of states $S_{a_1, \ldots,a_n,b}$.\footnote{The operation $p_a \otimes p_b$, where $p_a = \sum_{i \in I} y^{p_a[i]}$ and $p_b = \sum_{j \in J} y^{p_b[j]}$, is such that for each problem $(i,j) \in I \times J$ yields the solutions to $i$ and $j$, $p_a[i]$ and $p_b[j]$.}
\item Each object $a$ has an {\em identity} morphism.
\item Pairs of morphisms compose.
\end{itemize}

\noindent The last two requirements indicate, roughly, that morphisms inherit the identity and compositionality properties of $\mathbf{\mathbb{O}rg}$\\
.
\begin{teo} Both $Ob(\mathcal{G}) \subseteq Ob(\mathbf{\mathcal{PR - G}})$ and $Ob(\mathcal{PR}) \subseteq Ob(\mathbf{\mathcal{PR - G}})$.  
\end{teo}

\noindent {\bf Proof}: {\it Each problem in $\mathcal{PR}$ can be interpreted as an interface between the problem itself and its optimal solutions. The same applies to any interactive decision-making setting in $\mathcal{G}$.\\

\noindent More precisely, a local problem $s^k \in Ob(\mathcal{PR})$ and a game $G \in Ob(\langle \mathcal{G}, Eq\rangle)$ can be represented by polynomial functor $p_{s^k}$ or $p_G$, respectively. In the former case, $p_{s^k}$ is an interface between the specification of the local problem $(\hat{L}^k, u^k)$ and its solutions $\hat{\mathbf{X}}^k$. In the case of a game, $p_G$ is an interface between the game $G$ and its equilibria $Eq(G)$.}\\

\noindent {\it Each state in the morphism between two interfaces $p_{s^k}$ and $p_{s^j}$ represents a particular $r^k_j: \Sigma(s^k) \rightarrow \Sigma(s^j)$ that sends a section of solutions over $s^k$ to a corresponding section over $s^j$, yielding a {\em sheaf}. \\

\noindent Analogously, each state in the morphism between two interfaces $p_{G}$ and $p_{G^{\prime}}$ represents a particular {\em wiring}, connecting the games $G$ and $G^{\prime}$, such that the equilibrium obtains by tensoring those of the two games. $\Box$}\\

\noindent Notice that neither $\mathcal{PR}$ nor $\mathcal{G}$ are subcategories of $\mathcal{PR-G}$. While their objects are also objects of the latter, morphisms among them are not morphisms in $\mathcal{PR-G}$, which support dynamic rearrangements of the relations between its objects. Thus, $\mathcal{PR-G}$ incorporates all the representational advantages of $\mathcal{PR}$ and $\mathcal{G}$, adding the possibility of capturing the dynamics of actual systems.\\ 

\noindent The following two examples exhibit the representational power of $\mathcal{PR-G}$: 

\begin{example} (\cite{Samuelson}): {\it Consider a Principal-Agent problem defined by two functions:
$$\Phi_{\rightarrow}: X \times Y \times \mathbb{R} \rightarrow \mathbb{R} \ \mbox{and} \ \Pi: X \times Y \times \mathbb{R} \rightarrow \mathbb{R}$$
\noindent where:
\begin{itemize}
\item $X$ is the compact set of types of the Agent.
\item $Y$ is the compact set of possible decisions made by the Agent.
\item $\Phi_{\rightarrow}$ is continuous, strictly decreasing in the third argument.
\item $\Phi_{\rightarrow}$ is full range in the third argument: $\Phi_{\rightarrow}(x,y, \cdot)[\mathbb{R}] = \mathbb{R}$ for every $(x,y) \in X \times Y$.
\item $\Pi$ is continuous and increasing in the third argument.
\item $\Pi$ is full range in the third argument: $\Pi(x,y, \cdot)[\mathbb{R}] = \mathbb{R}$ for every $(x,y) \in X \times Y$.
\end{itemize}

\noindent Given a type $x$ of the Agent, her decision $y$ and $v$, the money transfer to the Principal, $\Phi_{\rightarrow}(x,y, v) = u_A$ is the utility of the Agent, while $\Pi(x,y, v) = u_P$ is the utility of the Principal.\\

\noindent An {\em inverse generating function} is

$$\Phi^{\leftarrow}: Y \times X \times \mathbb{R} \rightarrow \mathbb{R}$$

\noindent such that given $u_A = \Phi_{\rightarrow}(x, y, \Phi^{\leftarrow}(y, x, u_A))$ there exists $v = \Phi^{\leftarrow}(y, x, \Phi_{\rightarrow}(x, y, v))$.\\

\noindent Given $\lambda \in \mathbb{M}$, the class of {\em Borel measures} over $X \times Y$ and $\underline{u}$, a reservation utility of the Agent, the Principal's problem amounts to choosing $\langle \lambda, \bar{u}_A, \bar{v}\rangle$ as to maximize

$$ \int_X \int_Y \Pi(x,y, \Phi^{\leftarrow}(y, x, \bar{u}_A)) d \lambda(x,y)$$
\noindent s.t. $\bar{v} = \Phi^{\leftarrow}(y, x, \bar{u}_A)$ and $\bar{u}_A \geq \underline{u}$.

\noindent This setting can be naturally represented by defining two objects in $\mathcal{PR-G}$, $A$, and $P$ (the Agent and the Principal, respectively). The corresponding polynomial functors are:
\begin{itemize}
\item $p_P$ takes as input $\underline{u}$ and returns the optimal values $\lambda^*$, $u^*_A$ and $\bar{v}^*$. That is, $p_P = \sum_{\underline{u} \in \mathbb{R}} y^{p_P[\underline{u}]}$, such that $p_P[\underline{u}] = \langle \lambda^*, u^*_A, \bar{v}^* \rangle$.
\item $p_A$ takes as input $\bar{v}$ and returns her decision $y$ and the Principal's utility $u_P$. That is, $p_A = \sum_{\bar{v} \in \mathbb{R}} y^{p_A[\bar{v}]}$, such that $p_A[\bar{v}] = \langle y, u_P\rangle$.
\end{itemize}
\noindent Then, the entire problem can be understood in terms of the {\em identity} morphism of $p_A \otimes p_P$, yielding the adjunction between $\Phi^{\rightarrow}$ and $\Phi^{\leftarrow}$.
}
\end{example}

\noindent A promising area of research in which $\mathcal{PR-G}$ could be relevant for the design of mechanisms:

\begin{example} (\cite{Hurwicz} \cite{Jules}): {\it Mechanisms\footnote{Institutions as well.} can be conceived as game forms. That is, each mechanism $M$ can be represented as $M=(I_M, S_M, \mathbf{O}_M, \rho_M)$ (see Section~4).\\

\noindent Each $i \in I_M$ can be given different incentives according the environment $\mathbf{e} \in E$ in which she interacts with the others. Each $\mathbf{e} \in E$ will have an associated profile of payoff functions that correspond to the outcomes in $M$, $\pi_{M}^{\mathbf{e}}$.\\

\noindent The task of a mechanism designer $D$ is to assign to a given environment a mechanism $M \in \mathbb{M}$, in order to ensure a target $\mathbf{o}^*$. Thus, in $\mathcal{PR-G}$, $D$ has an associated $p_D = \sum_{\mathbf{e} \in E} y^{p_D[\mathbf{e}]}$ where $$p_D[\mathbf{e}] = \{\langle M, \pi_{M}^{\mathbf{e}} \rangle: M \in \mathbb{M} \ \mbox{such that} \ s^{*}_{M}  \in \mathbf{\mbox{Eq}}(\langle M, \pi_{M}^{\mathbf{e}} \rangle) \ \mbox{and} \ \rho(s^{*}_M) = \mathbf{o}^{*} \in \mathbf{O}_M\}$$

\noindent Each game form $M \in \mathbb{M}$ constitutes a {\em local problem}. The polynomial corresponding to these problems is $p_{\mathbb{M}}$. In turn, given the choice of {\em Nature} (represented by a constant polynomial $p_{E} = E$), the whole problem can be described by a $[p_D \times p_E, p_{\mathbb{M}}]$-coalgebra, where:

$$[p_D \times p_E, p_{\mathbb{M}}] \ = \ \sum_{\phi: p_D \times p_E \rightarrow p_{\mathbb{M}}} y^{\sum_{\mathbf{e} \in E} p_{\mathbb{M}}[\phi^{\rightarrow}(\mathbf{e})]}$$

\noindent and $p_{\mathbb{M}}[\phi^{\rightarrow}(\mathbf{e})] = \langle M, \pi^{\mathbf{e}}_M \rangle$.
}
\end{example}

\section{Conclusions}

\noindent This paper discussed the question of representing economic phenomena in terms of interactions among intentional agents. We resorted to the language of Category Theory and, in particular, constructions like {\em sheaves}, {\em hypergraph categories}, and {\em polynomial functors}.\\

\noindent The category defined in terms of the latter, $\mathcal{PR-G}$, has as objects the interfaces between problems and their solutions, while the interaction among them is captured by coalgebras based on the internal homs of the interfaces. That is, sets of states that determine the arrangement of connections among the problems and their solutions. Furthermore, the connections are rearranged in response to the outputs obtained previously.\\

\noindent We intend to explore further this formalism and use it to represent specific economic problems. While a first step involves showing that $\mathcal{PR-G}$ can reformulate known models, the real gist of this development is to capture new phenomena, establishing their relations to the former. \\

\end{document}